\documentclass[useAMS,usenatbib]{mn2e}

\usepackage{graphicx}
\newcommand{\um}{\hbox{\,$\mu$m}}
\newcommand{\hr}{\hbox{$^\mathrm{h}$}}
\def\min{\hbox{$^\mathrm{m}$}}
\def\sec{\hbox{$^\mathrm{s}$}}
\newcommand{\kms}{\hbox{\,km s$^{-1}$}}
\def\h2{\hbox{H$_2$}}
\def\msun{\hbox{M$_\odot$}}
\def\apj{\hbox{ApJ}}
\def\apjl{\hbox{ApJL}}
\def\apjs{\hbox{ApJS}}
\def\aj{\hbox{AJ}}
\def\araa{\hbox{ARA\&A}}
\def\mnras{\hbox{MNRAS}}
\def\aap{\hbox{A\&A}}
\def\rmxaa{\hbox{\emph Rev. Mex. Astr. Astrofys.}}

\usepackage{color}
\definecolor{grass}{rgb}{0,0.5,0}

\title[The NGC\,2264\,G jet]{\emph{Spitzer} imaging of the jet driving the NGC\,2264\,G outflow}
\author[P. S.  Teixeira et al.]{P. S. Teixeira$^{1,2,3}$\thanks{E-mail:
pteixeira@cfa.harvard.edu }, C. M$^{\textrm c}$Coey$^{4}$\thanks{E-mail: cmccoey@astro.uwaterloo.ca}, M. Fich$^4$\thanks{E-mail: fich@astro.uwaterloo.ca} and C. J. Lada$^1$\thanks{E-mail: clada@cfa.harvard.edu}\\
$^{1}$Harvard-Smithsonian Center for Astrophysics, 60 Garden Street,
 Cambridge, MA, 02138\\
$^{2}$Departamento de F\'{\i}sica da Faculdade de Ci\^encias da \hbox{Universidade} de Lisboa, Ed. C8, Campo Grande, 1749-016, \hbox{Lisboa},Portugal\\
$^3$Laborat\'orio Associado Instituto D. Luiz - SIM, Universidade de Lisboa, Campo Grande, 1749-016, \hbox{Lisboa}, Portugal\\
$^4$Department of Physics \& Astronomy, University of Waterloo, Waterloo, ON N2L 3G1, Canada}
\begin{document}

\date{}

\pagerange{\pageref{firstpage}--\pageref{lastpage}} \pubyear{2007}

\maketitle

\label{firstpage}

\begin{abstract}
We present new infrared imaging of the NGC\,2264\,G protostellar outflow region, obtained with the InfraRed Array Camera (IRAC) on-board the \emph{Spitzer} Space Telescope. A jet in the red outflow lobe (eastern lobe) is clearly detected in all four IRAC bands and, for the first time, is shown to continuously extend over the entire length of the red outflow lobe traced by CO observations. The redshifted jet also extends to a deeply embedded Class\,0 source, VLA\,2, confirming previous suggestions that it is the driving source of the outflow \citep{gomez94}. The images show that the easternmost part of the redshifted jet exhibits what appear to be multiple changes of direction. To understand the redshifted jet morphology we explore several mechanisms that could generate such apparent changes of direction. From this analysis, we conclude that the redshifted jet structure and morphology visible in the IRAC images can be largely, although not entirely, explained by a slowly precessing jet (period $\approx$ 8000\,yr) that lies mostly on the plane of the sky. It appears that the observed changes in the redshifted jet direction may be sufficient to account for a significant fraction of the broadening of the outflow lobe observed in the CO emission.
\end{abstract}


\section{Introduction}
\label{sec:introd}
It is well established that the early stages of star formation are accompanied by an energetic molecular outflow that is driven by a fast underlying wind. In contrast, the nature of this wind remains disputed and is typically modelled as either a wide-angle wind \citep[e.g.][]{shu91,shu95,lee01} or a collimated jet \citep[e.g.][]{raga93,masson93,stahler94,konigl00}. The latter class of model is made particularly appealing by the observation that protostellar jets and molecular outflows are often associated with the same young stellar objects \citep[e.g.][]{bachiller96}. However, models of jet-driven outflows are unable to produce the wide cavities often displayed in CO maps. It has frequently been proposed that variation in the direction of the jet propagation may act to broaden the outflow \citep{raga93}. Indeed, there is mounting observational evidence that jets do change direction within their outflows \citep[e.g.][]{bence96,schwartz99,davis00,arce01,arce02}. Along with the possibility that jets can act to broaden outflows, there is recent evidence of a precessing jet excavating a protostellar envelope \citep{ybarra06}. Nevertheless, clear evidence that a jet can broaden a molecular outflow to an extent comparable to that seen in the CO observations is still lacking. 

\citet{fich97} found indications of jet induced broadening of an outflow in the case of NGC\,2264\,G. This is a powerful, early-stage, bipolar molecular outflow discovered in a survey of high-velocity gas in NGC\,2264 \citep{margulis86}, situated at a distance of 800\,pc \citep{walker56}. \citet{gomez94} suggested that the driving engine for this outflow is VLA\,2, a protostar that is located between the two CO lobes. These researchers found that this radio continuum source, observed at 6\,cm and 3.6\,cm, is spatially coincident with an ammonia clump (T $\approx$ 15\,K). They additionally found that the source observed at 3.6\,cm presented an elongated axis that was approximately parallel to that of the NGC\,2264\,G outflow and proposed that the continuum emission correspond\textcolor{blue}{s} to a collimated jet. \citet{ward-thompson95} observed VLA\,2 with the JCMT and classified it as a Class\,0 protostar based on its ratio of submillimeter (L$_\mathrm{submm} \approx$0.3\,L$_\odot$) and bolometric luminosities (L$_* \approx$12\,L$_\odot$). They estimate a mass for the dust continuum source (VLA\,2+envelope) of \hbox{2-4\msun}.

\begin{figure*}
\centering
\includegraphics[width=\textwidth]{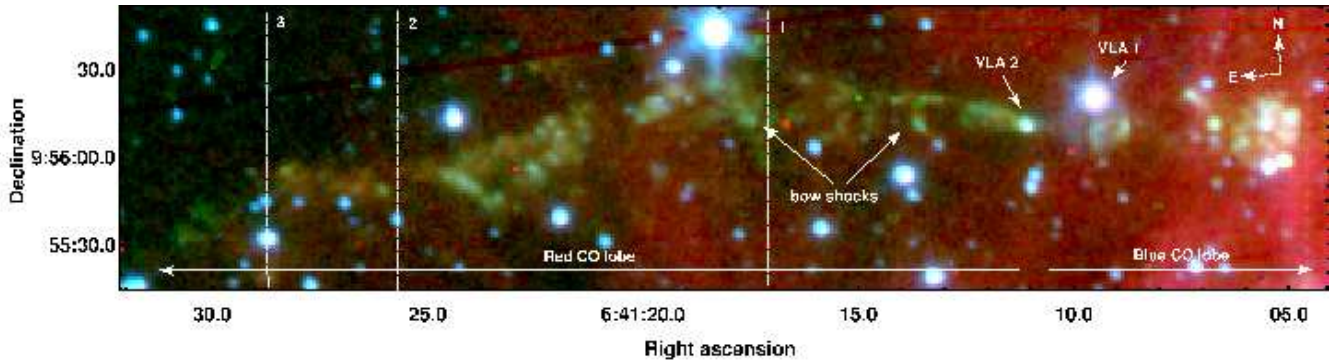}
\caption{Colour composite image of the NGC\,2264\,G outflow region, where red corresponds to 8\um, green to 4.5\um, and blue to 3.6\um\ IRAC data. We see the jet associated with the red lobe of the outflow in greenish hues, extending over 1.1\,pc \citep[assuming a distance of 800\,pc;][]{walker56}. The numbered dashed vertical lines mark the locations where the jet appears to change direction in the red CO lobe (coordinates $(\alpha_1,\delta_1),\ (\alpha_2,\delta_2)$, and $(\alpha_3,\delta_3)$, see text for details), while the bottom horizontal arrows indicate the location of the red and blue CO emission lobes \citep{margulis86}. The driving source of the jet, VLA\,2, is identified as well as VLA\,1 and bow shock structures of the jet.}
\label{fig:IRACcolour}
\end{figure*}

NGC\,2264\,G extends $\approx$ 1.6\,pc along an east-west direction \citep{CoG06}, is highly collimated, has a mechanical luminosity of $\approx$1\,L$_\odot$ and a mass of $\approx$1\msun\ \citep{lada96}, and has been extensively mapped in lines of CO \citep[e.g.][]{margulis86,fich98}. The outflow displays a ``Hubble law'' velocity distribution, from which \citet{lada96} were able to estimate an upper limit to the dynamical time of 3-4$\times10^4$\,yr.  In addition, there is an `\emph{S}-shaped' symmetry to the outflow, with low velocity ($\approx$ 6\kms) CO emission tracing a cavity near to the source while the more collimated high velocity ($\approx$ 40\kms) CO emission lies along a different axis toward the end of the outflow \citep{lada96}.  Finally, the high degree of collimation and highly bipolar nature of the CO outflow lead \citet{fich97} to conclude that the outflow must lie within 20 degrees of the sky.  

NGC\,2264\,G has also been imaged in \h2\ v=1-0 S(1) by \citet{davis95}, who interpreted bright knots at the east edge of the outflow as a signature of a turbulent mixing layer at the boundary between outflow and ambient medium, and at higher resolution by \citet{CoG06}, who found chains of knots they attribute to a jet along the outflow. 
 \citet{fich97} compared the CO maps of \citet{lada96} with the \h2\ images of \citet{davis95} and noted that the bright \h2\ knots coincided with points at which the axis of CO changed direction. On the basis of this, they suggested that the high velocity CO emission in both lobes of the outflow traced the molecular component of a deflected jet. Furthermore, they speculated that such deflections could serve to broaden the CO outflow lobes to an extent comparable to that observed. The aim of this paper is to investigate their hypothesis.
 
The \emph{Spitzer} Space Telescope has provided the astronomical community with striking images of protostellar jets \citep[e.g][]{noriega04}. The plethora of molecular lines that fall within the InfraRed Array Camera (IRAC) broad band filters, combined with the high sensitivity of the detector, show detailed features that are observationally unaccessible to near-infrared or optical imaging, or even to ground-based mid-infrared imaging. Here, we present IRAC images of the NGC\,2264\,G outflow that clearly show a jet, which appears to exhibit three changes of direction in its easternmost part.  In conjunction with CO maps from \citet{fich98}, we reconsider the evidence for deflection of the outflow presented by \citet{fich97} and explore other jet-bending mechanisms, in particular, precession. 

The structure of the paper is the following: in \S\,\ref{sec:obs_red} we give a brief description of the observations and data reduction; in \S\,\ref{sec:res} we present the results obtained from our analysis; and, in \S\,\ref{sec:dis} we discuss several possible jet bending mechanisms with a view to explaining the observed jet morphology. Finally, in \S\,\ref{sec:concl} we summarize our main conclusions.

\section{Observations and reduction}
\label{sec:obs_red}

NGC\,2264\,G was observed with IRAC onboard the \emph{Spitzer} Space Telescope as part of the \emph{Spitzer} Guaranteed Time Observation program 37 \citep{fazio04, 06}. The data were acquired in two epochs, with two dithers each, and consist of observations in all four IRAC bands (3.6, 4.5, 5.8, and 8\um). Each dither observed consists of consecutive exposures of 0.4\,s and 10.4\,s integration times. Data reduction and calibration were performed with the \emph{Spitzer} Science Center (SSC) pipeline, version S14. We combined the four images downloaded from the SSC archive into a final mosaic for each band.

\section{Results}
\label{sec:res}

\begin{figure*}
\centering
\includegraphics[width=\textwidth]{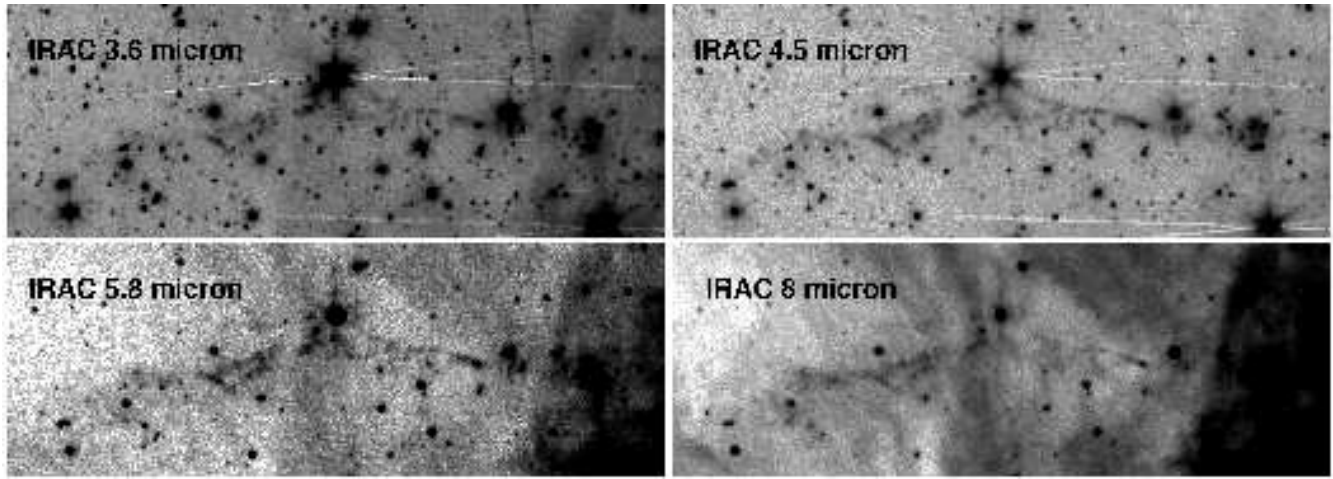}
\caption{IRAC greyscale images of the NGC2264\,G region.  The jet is detected in all four IRAC bands but is brightest at 4.5\um. }
\label{fig:IRACbands}
\end{figure*}

Figure \ref{fig:IRACcolour} shows a colour composite image of the red lobe (east) and part of the blue lobe (west) of the NGC\,2264\,G outflow region, built from 3.6, 4.5, and 8\um\ data. We detect emission from a jet in all four IRAC bands, as shown in Figure \ref{fig:IRACbands}; the jet is brightest in band 2 (4.5\um) and takes the form of a narrow, near-continuous feature that extends over the entire length of the red lobe (1.1\,pc) of the CO outflow. As we can see from Figure \ref{fig:IRACcolour}, the redshifted jet clearly traces back to the protostar VLA\,2, directly confirming that this source is indeed driving the jet and outflow, as previously suggested by \citet{gomez94}. We do not detect a jet corresponding to the blue-shifted CO lobe; instead, we observe only a tight clustering of bright knots ($\alpha \approx$ 06\hr41\min06\sec, 06\hr41\min08.5\sec). In general, one expects the redshifted jet, and its associated outflow lobe, to move deeper into the parental molecular cloud and hence suffer more extinction than the blue counterpart; this is not the case for NGC\,2264\,G. 

We note that the IRAC 5.8\um\ and 8\um\ broad bands include Polycyclic Aromatic Hydrocarbon (PAH) emission lines (6.2\um\ and 7.7-7.9\um) \citep{noriega04}. There is increasingly bright and extended PAH emission detected in the 8\um\ IRAC band (red) towards the west (see bottom panels of Figure \ref{fig:IRACbands}), indicating the presence of the molecular cloud.  This PAH emission likely arises due to the UV radiation field of the nearby O7 star, S\,Mon, \citep[($\alpha,\delta$)(J2000)=(06\hr40\min58.7\sec, +09\degr53\arcmin45\arcsec)][]{herrero92}. 

It is clear from Figure \ref{fig:IRACcolour} that VLA\,2 is situated towards the edge of the molecular cloud. As a result, the red lobe (and redshifted jet) travels into the less dense ambient medium, while the blue lobe (and blueshifted jet) push more deeply into the molecular cloud and is consequently highly extincted and mostly hidden. Therefore, we restrict our analysis to the jet in the redshifted CO lobe of the NGC\,2264\,G outflow. 

The colour image also shows what appear to be several changes of direction of the redshifted jet. We note that the main change of direction is seen in the \h2\ image of \citet{CoG06}. The IRAC broad bands (particularly the 4.5\um\ band) include many \h2\ quadrupole transition lines \citep[e.g.][]{smith05} and other molecular lines that are associated with low-mass outflows, such as CO vibrational lines and lines of PAHs \citep[e.g.][]{lefloch03}. The increased sensitivity of \emph{Spitzer}'s detectors relative to those of the same wavelength on ground-based telescopes, combined with the wealth of \h2\ lines covered by the IRAC bands, results in a much clearer and more complete image of the redshifted jet.

Figure \ref{fig:IRACcolour} presents 
the locations where the redshifted jet seems to change direction. Before the first change of direction, at \hbox{$(\alpha_1, \delta_1)$(J2000)=(06\hr41\min17.0\sec, +09\degr56\arcmin20\arcsec)}, we see several bow shock-shaped structures, which may indicate internal shocks along the redshifted jet. The redshifted jet subsequently seems to broaden before changing direction twice more at \hbox{$(\alpha_2, \delta_2)$(J2000)=(06\hr41\min25.5\sec, +09\degr55\arcmin45\arcsec)} and finally at \hbox{$(\alpha_3, \delta_3)$(J2000)=(06\hr41\min28.5\sec, +09\degr55\arcmin52\arcsec)}. The remainder of the eastern part of the redshifted jet is fainter, but still visible (SNR $\approx$ 2 at 4.5\um).

\section{Discussion}
\label{sec:dis}

\subsection{Origin of jet morphology}
\label{subsec:mechanisms}

Various mechanisms have been proposed to explain the bending structure of jets by a number of authors.  For example, \citet{eisloffel97} and \citet{fendt98} discuss the following possibilities that may cause apparent directional changes of the jet flow: (1) the impact due to side winds or overlapping outflows;  (2) Lorentz forces on a magnetic jet (coupling between a magnetic jet and an external magnetic field); (3) precession due to misalignment between rotation and magnetic fields; (4) deflection caused by density gradients in (or dynamical pressure of) the ambient medium; (5) motion of the powering source perpendicular to the outflow direction; and, (6) precession due to a binary system that has an orbital plane tilted with respect to the accretion disc. Mechanisms 1 and 5 produce `\emph{C}-shaped' jets, mechanisms 3 and 6 produce `\emph{S}-shaped' jets, while the remaining mechanisms (2 and 4) can produce jets with either \emph{C} or \emph{S} shapes.  

\begin{figure*}
\centering
\includegraphics[angle=90,width=\textwidth]{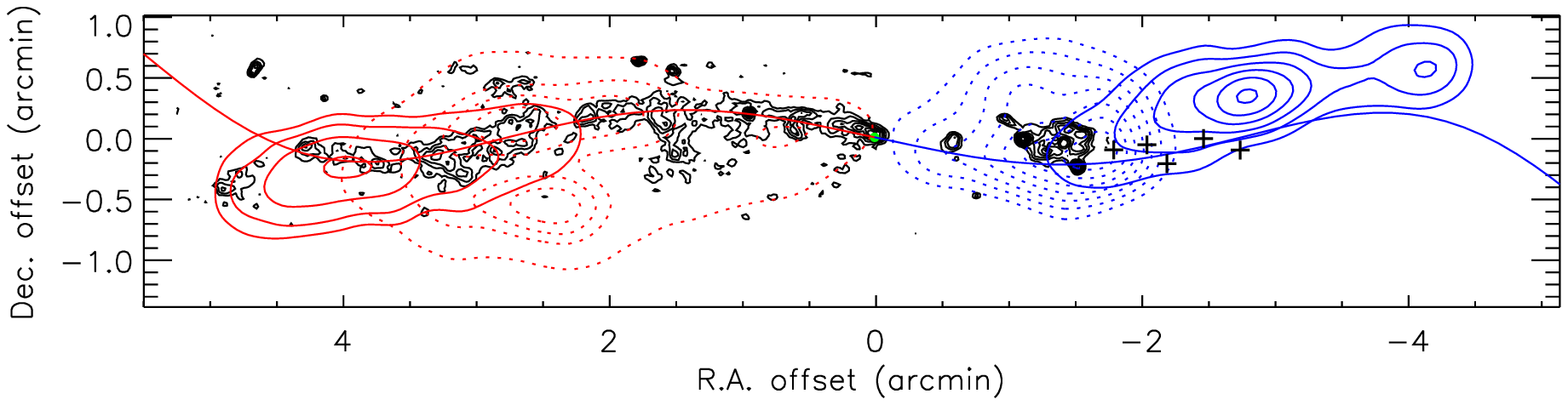}
\caption{We show the divided 4.5\um/3.6\um\ image in black contours (ranging from 1.25 to 3.1 in steps of 1$\sigma$, 0.15 MJy\,sr$^{-1}$), with CO emission (Fich \& Lada, 1998) from red and blueshifted lobes overplotted. The black plus signs mark the location of \h2\ knots identified by \citet{CoG06} and the green point represents the driving source, VLA2. The red solid and dotted contours correspond to high ($<$ 39\kms) and low ($<$ 8\kms) velocity redshifted CO, respectively. Likewise, the blue solid and dotted contours correspond to high ($<$ 27\kms) and low ($<$ 3\kms) velocity blueshifted CO, respectively. The solid red and blue lines are a precession model fit, as explained in the text.}
\label{fig:precession}
\end{figure*}

There is, to our knowledge, no published evidence for overlapping outflows in the NGC\,2264\,G region. Moreover, Figure\,3 of \citet{lada96} shows that the outflow is `\emph{S}-shaped' (the eastern red CO lobe curves in the opposite sense to the western blue CO lobe) and symmetric with respect to the driving source VLA\,2, meaning that the CO outflow does not form the characteristic overall \emph{C} shape or bending structure one would expect from side winds. Hence, mechanism 1 is effectively ruled out for NGC\,2264\,G. We show the CO data from \citet{lada96} in Figure \ref{fig:precession}, that will be discussed in more detail further below.

We are unable to discuss the plausibility of mechanisms 2 and 3 because, as yet, no information regarding the magnetic field of the region is available \citep[for description of these mechanisms see][]{eisloffel97,fendt98}.

We note there is a massive star near to NGC\,2264\,G ($\la$ 0.6\,pc southwest from the knots in the blue lobe), S\,Mon, \citep[($\alpha,\delta$)(J2000)=(06\hr40\min58.7\sec, +09\degr53\arcmin45\arcsec), spectral type O7\,Ve;][]{herrero92} whose wind and radiation pressure could act to force the outflow northward.  Indeed, \citet{fich97} found that the blue lobe of the outflow does bend northward: Figure 1 of their paper shows that the axes of the low and high velocity CO change their position angles at a location coincident with \h2\ knots \citep[][also observed in our IRAC images, R.A.$\approx$06\hr41\min05\sec, see Fig \ref{fig:precession}]{davis95}. However, if this change in direction was caused by S\,Mon alone we would expect it to occur further to the west, closer to the massive star. In \S\,\ref{subsubsec:disCO} we will discuss further how mechanism 4, in particular the molecular cloud structure, could be shaping the NGC\,2264\,G outflow. 

Mechanisms 5 and 6 can occur in binary systems if the binary orbital plane and accretion disc of the star powering the outflow are coplanar (5) or non-coplanar (6). The 5th mechanism can additionally occur if the powering source is ejected from its birthplace \citep{goodman04}.
 In the case of mechanism 5, the proper motion of the powering source is perpendicular to the jet axis, giving rise to an asymmetric (`\emph{C}-shaped') jet morphology \citep[for a detailed explanation, see][]{fendt98}. As noted above, NGC\,2264\,G is a symmetric outflow so this mechanism can also be ruled out.   On the other hand, a binary system with an orbital plane that is non-coplanar with the disc of the powering source can induce wobbling or precession of the accretion disc and subsequently cause precession of the jet \citep{terquem99,bate00}. We will discuss this 6th mechanism in more detail below, in \S\,\ref{subsec:precess}.

\subsubsection{The deflection scenario}
\label{subsubsec:disCO}

It was suggested by \citet{fich97} that the NGC\,2264\,G outflow is shaped by a jet that is deflected off the cavity wall of the outflow. This hypothesis was based on the fact that \cite{davis95} detected H$_2$ knots in the regions where the axis of the red CO lobe was found by \citet{lada96} to change direction. Indeed, we note that the path of the redshifted jet and position of the first deflection point are in excellent agreement with the predictions of \citet{fich97}.

In this scenario, the redshifted jet initially flows along the centre of the red lobe of the outflow, between the two low-velocity CO peaks seen in Figure \ref{fig:precession}. A change in the angle at which the driving source ejects material then results in the redshifted jet flowing along the northern edge of the low velocity CO emission, before deflecting off the inner cavity wall approximately halfway along the outflow and continuing south-eastward. One would expect the jet to fan out after interacting with a denser medium such as a cavity wall or a (clump in the) molecular cloud \citep{raga95,davis00} and there is some trace of this behavior in Figure \ref{fig:IRACcolour} after the first change in direction.

This change of position angle of the jet at the driving source could have been a single abrupt change, a series of abrupt changes, or a continuous change.  
The latter type of change could be due to precession as we discuss in the next section, \S\,\ref{subsec:precess}. An abrupt change in the ejection angle of the jet could be provoked by close interaction with another stellar member of NGC\,2264, or ejection events. We have no evidence either to support or rule out the possibility that an abrupt change (or changes) did indeed occur.
 
The IRAC image shows us that the redshifted jet appears to undergo two further directional changes, one at the point at which the redshifted jet pierces the outflow cavity (denoted by the easternmost edge of the low velocity CO emission \hbox{$(\alpha_2, \delta_2)$(J2000)=(06\hr41\min25.5\sec, +09\degr55\arcmin45\arcsec)}) and another further downstream (\hbox{$(\alpha_3, \delta_3)$(J2000)=(06\hr41\min28.5\sec, +09\degr55\arcmin52\arcsec)}).  Although the deflection scenario has some appealing aspects, the changes of direction in the red lobe require several deflections of the redshifted jet off the ambient material and this, while possible, seems rather unlikely.

\subsubsection{The precession scenario}
\label{subsec:precess}

\begin{figure}
\centering
\includegraphics[width=0.35\textwidth]{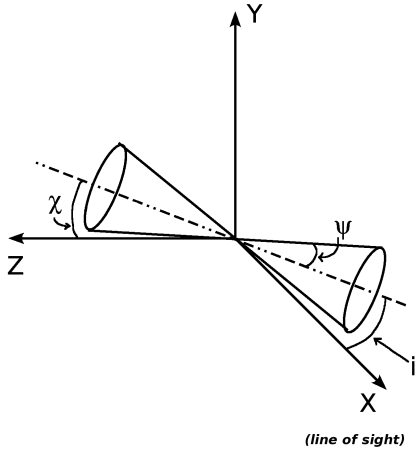}
\caption{Schematic diagram depicting the orientation of the jet. The line of sight is along the X axis and the plane of the sky is YZ. The inclination angle \emph{i} of the jet axis (dashed-dotted line) with respect to the line of sight to the observer is measured in the plane XZ, while $\chi$ is the angle of the jet axis in the plane of the sky. The jet is rotating within a cone that is characterized by an opening angle 2$\psi$.}
\label{fig:diagram}
\end{figure}

In this section we explore the possibility that a continuous, rather than abrupt, change of jet direction -- precession -- could explain the morphology of the jet in the red lobe. This scenario has been used to describe the structure of other jets driven by protostellar sources, in particular Cep R \citep{eisloffel96}.  As mentioned in \S\,\ref{subsec:mechanisms}, a binary system in which the binary orbital plane and the accretion disc of the powering source are non-coplanar can lead to precession of the jet, which has a time-scale that increases with binary separation \citep{terquem99, bate00}.  Precession could also be caused by warping instabilities in an accretion disc that is threaded by a large-scale magnetic field \citep{lai03}. Could precession fully explain the morphology of the NGC\,2264\,G redshifted jet without invoking a deflection mechanism? To attempt to answer this question we use a simple nonrelativistic jet model from \citet{hjellming81} and \citet{gower82}, which is described by the following equations:

\begin{displaymath}
\left[ \begin{array}{cc}
\alpha   \\
\delta  \end{array} \right] =
 \left[ \begin{array}{cc}
\alpha_0   \\
\delta_0  \end{array} \right]
+
 \left[ \begin{array}{cc}
sin(\chi) & cos(\chi)   \\
cos(\chi) & -sin(\chi)  \end{array} \right] 
\,s_{jet}\,\times
\end{displaymath}
\begin{displaymath}
\times \left[ \begin{array}{cc}
s_{rot} sin(\psi) sin(2\pi l/\lambda + \phi_0)   \\
s_{rot} cos(\psi) sin(i)-sin(\psi) cos(i)cos(2\pi l/\lambda + \phi_0)  \end{array} \right] 
\end{displaymath}

\noindent
where $(\alpha_0,\delta_0)$ are the coordinates of the driving source VLA\,2; $l$ is the distance between $(\alpha,\delta)$ and $(\alpha_0,\delta_0)$; $\phi_0$ is the initial phase of the precession at the source; $\lambda$ is the precession length scale; $s_{jet}$ is the sign for the jet (-1) or counterjet (1); and, $s_{rot}$ is the sense of rotation (-1 for clockwise and 1 for anti-clockwise rotation). Figure \ref{fig:diagram} shows $\psi$, $\chi$, and $i$ which are the precession angle, the angle of the jet axis in the plane of the sky, and the inclination angle of the blueshifted lobe to the line of sight, respectively. We note that cognizance of the inclination angle of the jet axis to the line of sight  is very important in order to estimate the momentum and energy of the jet. Figure \ref{fig:precession} shows a precession model overplotted on a divided 4.5\um/3.6\um\ image that has the following parameters (for a anti-clockwise rotating jet): $\chi$=-185\degr, $\phi_0$=75\degr, $\psi$=8\degr, $i$=82\degr, and $\lambda$=6.9\arcmin. For a anti-clockwise rotating jet we find that $\lambda$ may vary between 6.5\arcmin\ and 7.5\arcmin\ and $i$ may vary between 70\degr\ and 82\degr. For a clockwise rotating jet we find that the model describes the blueshifted lobe better but does not follow the redshifted lobe so well for the same group of parameter values. Assuming a typical velocity for jets driven by young protostars of 200\kms\ \citep[e.g.][]{eisloffel96,gomez97} we estimate that the precession period is 7300-8500\,yr for a distance of 800\,pc to NGC\,2264\,G. \citet{lada96} found an upper limit for the dynamical time-scale of the gas in the outflow to be on the order of 10$^4$\,yr, with which this precession scenario is consistent.

Assuming the high velocity CO traces a molecular component of the jet, we may infer that the radial, or line of sight, velocity of the jet is $\approx$ 40\kms. For a total jet velocity of 200\kms\ this corresponds to an inclination angle to the line of sight $i \ga$ 78\degr\ (i.e., within 22\degr\ from the plane of the sky).  \citet{lada96} found the outflow to be highly bipolar and concluded that the velocity vectors of the outflowing gas are directed predominately along the axis of the flow.  From this and the high degree of collimation of the outflow, they concluded that the opening angle of the flow measured from the jet axis, $\psi$, be 8\degr\ and that the outflow should lie within 20\degr\ of the sky. Both of these constraints are consistent with our precession model.

We note that the small angle to which the jet is inclined with respect to the plane of the sky  requires high resolution spectroscopic observations to measure the radial velocity of the jet. Unfortunately, ground-based high resolution mid-IR spectral observations are not currently feasible because the jet is too faint and present space-based spectrometers lack the required resolution. We compared our IRAC images and the \h2\ images of \citet{davis95} and found no discernible proper motion of the knots along the redshifted jet. Future imaging observations may yield proper motion measurements and provide a better estimate of the precession time-scale using the precession length scale we determined via our model. Nonetheless, the inclination angle corresponding to a 200\kms\ jet ($i \ga$ 78\degr) is consistent with what we derive from our model (70\degr $\le i \le$ 82\degr).

If the precession is caused by a binary companion, and we take the precession period to be equal to the binary period (assuming 1\msun\ for VLA\,2), then the binary separation would be roughly 400\,AU. Alternatively, if we follow \citet{bate00} and assume the orbital period of the binary is 20 times that of the precession period then the binary separation would be about 3000\,AU. In either case, we are unable to resolve this hypothetical binary with the present data. \citet{gomez94} measured an ammonia condensation centred on VLA\,2 with a size of 0.08\,pc$\times$0.06\,pc ($\approx$16,000$\times$12,000\,AU). Assuming the ammonia emission arises from a circumstellar (or circumbinary) envelope then the hypothetical binary cannot be ruled out since its separation is within the envelope size. 

As can be seen by Figure \ref{fig:precession}, our precession model does not explain the easternmost knots that lie within the high velocity CO emission, although it fits the rest of the emission detected by IRAC in the red CO lobe very well. Even though the precession model seems to follow the \h2\ knots detected by \citet{CoG06} in the region of the blue lobe, it does however fail to properly describe the lobe's morphology as seen by the CO emission because the blue lobe is displaced further to the north relative to the path predicted by the precession model.
To account for this displacement we may need to additionally invoke a deflection similar to that suggested by \citet{fich97}.

\subsubsection{Jet-induced broadening}
\label{subsubsec:broadening}

\citet{arce06} found a linear correlation between the opening angle of protostellar outflows and the evolutionary stage of the associated powering source. According to their result, one would expect a small opening angle for an outflow driven by a Class\,0 source such as VLA\,2. This is in fact the case: \citet{lada96} measured an opening angle for the low velocity CO of $\approx$ 25\degr. How does the opening angle of the molecular outflow relate to that of the jet? 
As mentioned in Section \S\ref{sec:introd}, it remains observationally unclear as to what extent a jet may actually broaden an outflow. The ``sideways splash'' of material at the head of bow shocks is often cited as a means of broadening a jet-driven outflow \citep{raga93}. We note the presence of bow shock-shaped structures along the NGC\,2264\,G redshifted jet in Figure \ref{fig:IRACcolour} but the changes of redshifted jet direction are a much more significant feature. The interaction of the jet with ambient medium and cavity walls results in a transfer of momentum to the surrounding medium \citep{downes99}. If the jet has some sideways motion (due to precession or deflection) then the direction of momentum transfer can be perpendicular to the jet axis -- and so broaden the outflow. We find that the action of a precessing jet may further provide a significant contribution to the broadening of the NGC\,2264\,G outflow: our precession modeling gives a total opening angle of 16\degr, so this mechanism can account for roughly 64\% of the opening angle of the low velocity CO gas of NGC\,2264\,G.

\section{Summary and conclusions}
\label{sec:concl}

We analysed \emph{Spitzer} observations of the NGC\,2264 region with the aim of investigating the suggestion of \citet{fich97} that the NGC\,2264\,G outflow is shaped by a deflected redshifted jet. Here, we summarize our findings.
\begin{enumerate}
\item  A jet is clearly seen in all four \emph{Spitzer} IRAC images, in the red lobe of the NGC\,2264\,G outflow, and is brightest in the 4.5\,$\mu$m band. We find the redshifted jet to extend continuously across the entire length of the red outflow lobe. It appears to emanate directly from VLA\,2, confirming this Class\,0 source as the driving source of the outflow.
The most striking features of the redshifted jet image are three distinct points where the redshifted jet appears to change direction.
\item We discuss various mechanisms that can result in changes of jet direction and explore further the two most likely, deflection and precession. We are able to explain the majority of the redshifted jet structure, as seen by IRAC, with a simple precessing jet model. This model predicts a precession time of $\approx$8000\,yrs for a jet velocity of 200\kms. The precession of this jet may contribute to the broadening of the red lobe and account for its overall morphology, however, this model cannot satisfactorily explain the entire morphology of the blue CO lobe without invoking an additional change in direction or deflection of the underlying jet similar to that proposed by \citet{fich97}.
\end{enumerate}

\section*{Acknowledgments}

P.~S.~T. acknowledges support from the scholarship SFRH/BD/13984/2003 awarded by the Funda\c{c}\~ao para a Ci\^encia e Tecnologia (Portugal).
Both M.~F. and C.~M. are supported by a Natural Sciences and Engineering Research Council of Canada (NSERC) Discovery Grant.
This work is based [in part] on observations made with the Spitzer Space Telescope, which is operated by the Jet Propulsion Laboratory, California Institute of Technology under a contract with NASA.


\label{lastpage}

\end{document}